\documentclass[10pt, conference, compsocconf]{IEEEtran}
\usepackage{todonotes}
\usepackage{inconsolata}
\newcommand{\citep}[1] { \cite{#1}}
\newcommand{\citet}[1] { \cite{#1}}

\ifCLASSINFOpdf
\else
\fi

\usepackage[cmex10]{amsmath}
\begin{document}

\title{ALEA: Fine-grain Energy Profiling with Basic Block Sampling}

\author{\IEEEauthorblockN{Lev Mukhanov, Dimitrios~S. Nikolopoulos and Bronis R. de Supinski}
\IEEEauthorblockA{The School of Electronics, Electrical Engineering and Computer Science\\
Queen's University of Belfast\\
Belfast, UK\\
Email: \{l.mukhanov,dsn,b.de-supinski\}@qub.ac.uk}

}

\maketitle

\begin{abstract}
Energy efficiency is an essential requirement for all contemporary computing systems. 
We thus need tools to measure the energy consumption of computing systems and to understand how 
workloads affect it. Significant recent research effort has targeted direct 
power measurements on production computing systems using on-board sensors 
or external instruments. These direct methods have in turn guided studies of 
software techniques to reduce energy consumption via workload allocation and scaling. 
Unfortunately, direct energy measurements are hampered 
by the low power sampling frequency of power sensors. The coarse granularity of power
sensing limits our understanding of how power is allocated in systems and our ability to 
optimize energy efficiency via workload allocation. 

We present ALEA, a tool to measure power and energy consumption 
at the granularity of basic blocks, using a probabilistic approach. ALEA provides fine-grained energy 
profiling via statistical sampling, which overcomes the limitations of power
sensing instruments. Compared to state-of-the-art energy measurement tools, ALEA provides 
finer granularity without sacrificing accuracy. ALEA achieves low 
overhead energy measurements with mean error rates between 1.4\% and 3.5\% in 14 
sequential and parallel benchmarks tested on both Intel and ARM platforms. The sampling method caps execution 
time overhead at approximately 1\%. ALEA is thus suitable for online energy 
monitoring and optimization. Finally, ALEA is a user-space tool with a portable, machine-independent 
sampling method. We demonstrate three use cases of ALEA, where we reduce the energy consumption of 
a k-means computational kernel by 37\%, an ocean modeling
code by 33\%, and a ray tracing code by 6\% compared to high-performance execution baselines, by varying the power optimization 
strategy between basic blocks.
\end{abstract}


\IEEEpeerreviewmaketitle

\section{Introduction}
\label{sec:introduction}
Association of energy use with specific software abstractions and components 
enables the energy-efficient use of computing systems. Numerous energy profiling 
tools target platforms ranging from sensors, to smartphones, embedded systems, 
and high-end computing systems.

These tools guide software-controlled energy optimization techniques such as dynamic 
voltage and frequency scaling, thread packing, and concurrency throttling.

Emerging algorithmic energy models and metrics~\cite{Hsu:2012:TES:2188286.2188309,Alonso:2013:EED:2558756.2558785} 
for  high-level computation and communication abstractions make accurate energy 
accounting between software abstractions even more pressing. 

Prior energy accounting tools can be broadly classified into two categories: 
Tools that measure energy by directly measuring power using on-board 
sensors or external instruments~\cite{compaq_energy,powerscope99,PowerPack,energy_profiling_microsoft,Keranidis:2014:NMM:2602044.2602047,McIntire:2007:ESN:1236360.1236448};
and tools that model energy based on activity vectors of hardware  performance 
counters, kernel event counters, finite state machines, or instruction counters in 

microbenchmarks~\cite{Bertran:2013:SMG:2498752.2499068,Manousakis2014,prediction_profiling1,prediction_profiling2,dimitrios_bronis,Shao:2013:ECI:2648668.2648758,Tsoi:2011:PPO:2082156.2082159,Tu:2014:PPP:2597648.2566660,Wilke:2013:JGF:2451605.2451610,eprof_mobile,eprof}. 
All of these tools can associate energy measurements with software contexts via 
manual instrumentation, context tracing, or profiling. 

Energy accounting tools based on direct power measurement can accurately measure
both component-level and system-wide energy consumption, before and after the 
system's power supply units. However, the time granularity of the sensors 
fundamentally limits these tools. State-of-the-art external instruments such as 
the Monsoon power meter have sampling rates of at most 5 kHz~\cite{Brouwers:2014:NNE:2668332.2668337}. 
Some direct energy measurement and profiling tools use instruments with sampling 
rates as low as 1 Hz~\cite{PowerPack,powerscope99}. Internal energy and power 
sensors such as Intel's RAPL~\cite{intel_manual_rapl}
or the sensors commonly found on ARM-based boards~\cite{Cao:2012:YYP:2337159.2337185} have sampling 
frequencies between 1 and 3 kHz. The coarse granularity of direct power measurements 
limits their ability to account for the energy consumption of specific instructions 
or many software components such as basic blocks and most function instances, which 
typically execute for periods far shorter than the instrument sampling period.

Tools that model energy consumption from activity vectors can break the granularity 
barrier of direct energy measurements but suffer from several other shortcomings.
Their accuracy may be limited and highly dependent on architectural variations 
between platforms and workload patterns~\cite{prediction_profiling1,prediction_profiling2,dimitrios_bronis,eprof_mobile,eprof}. 
The tools require extensive training and benchmarking processes that must be 
repeated per platform and workload, to calibrate platform parameters.

This paper presents a new method that directly measures power consumption in 
computing systems and accounts for energy consumption of fine-grain code blocks,
including basic blocks with execution duration shorter than the minimum
power consumption sampling period. We use the term coarse-grain for basic blocks of longer 
duration. Our energy accounting tool combines the accuracy of direct power 
measurements with the fine granularity of energy accounting between basic blocks.
Our Abstraction-Level Energy Accounting (ALEA) tool uses the systematic sampling of 
physical power measurements and a probabilistic model to distribute energy between 
basic blocks of any granularity, while capturing the dynamic execution context of 
these blocks. ALEA achieves portability through a machine-independent sampling 
method that abstracts the details of the underlying architecture and power measurement
instruments. We demonstrate its accuracy, efficiency and portability on two 
multicore platforms based on the Xeon Sandy Bridge and Samsung Exynos processors. 
We validate ALEA with 14 sequential and parallel applications. 
ALEA's mean error for coarse-grain basic blocks, as well as for the whole program, 
is 1.4\% on the Sandy Bridge server and 1.9\% on the Exynos SoC. ALEA's mean 
error for fine-grain basic blocks is 1.6\% on the Sandy Bridge server and 3.5\% 
on the Exynos SoC. We use ALEA to demonstrate the correlation between power 
consumption and cache accesses at the basic block level across our benchmark suite. 
Finally, we demonstrate three use cases of ALEA, where we reduce the energy consumption of 
a k-means computational kernel by 37\%, an ocean modeling
code by 33\%, and a ray tracing code by 6\% compared to high-performance execution baselines, by varying the power optimization 
strategy between basic blocks. 
 
The rest of this paper is structured as follows. Section~\ref{sec:related_work} 
presents related work. Section~\ref{sec:platforms} describes our platforms and their 
direct energy measurement sensors. Section~\ref{sec:profiling} details our energy 
sampling and profiling models and the key aspects of their implementation. 
Section~\ref{sec:validation} validates ALEA's energy profiler. Section~\ref{sec:experiments} presents a use case
of ALEA in understanding the impact of memory accesses and thread synchronization on energy.
Section~\ref{sec:use_case} presents further use cases of ALEA for fine-grain energy optimization in parallel codes.
Section~\ref{sec:conclusion} summarizes our findings.

\section{Related Work}
\label{sec:related_work}

Statistical sampling of the execution context of a running program is an established 
method for performance profiling~\cite{hpctool_analysis,hpdctool_unwind,hpctoolkit}.
Sampling is also a state-of-the-art method for profiling large-scale data 
centers~\cite{gwp}. ALEA is the first tool to deploy basic block sampling and 
power sampling for fine-grain energy profiling. 

Several tools for energy profiling use manual instrumentation to collect samples 
of hardware event rates from hardware performance monitors (HPMs)~\cite{Bertran:2013:SMG:2498752.2499068,prediction_profiling1,prediction_profiling2,dimitrios_bronis,Li:2013:SER:2420628.2420808}. 
These tools empirically model power consumption as a function of one or more activity 
rates that attempt to capture the utilization and dynamic power consumption 
of specific hardware components. HPM-based tools and their models have guided several 
power-aware optimizations. However, they often estimate power with low accuracy.
Further, they rely on architecture-specific training and calibration.

PowerScope~\cite{powerscope99,powerscope_flinn}, an early energy profiling mechanism,
profiles mobile systems through direct hardware instrumentation. It samples power 
consumption, which it attributes to processes and procedures through post-processing. 
In contrast, ALEA profiles at a finer granularity. 
 
Eprof~\cite{eprof_mobile,eprof} models hardware components as finite state machines
with discrete power states and emulates their transitions to attribute energy use
to system calls. 
JouleUnit~\cite{Wilke:2013:JGF:2451605.2451610} correlates workload profiles with 
external power measurements to derive energy profiles across method calls.  
JouleMeter~\cite{energy_profiling_microsoft} uses post-execution event tracing to
map measured energy consumption to threads or processes.
These tools perform energy accounting at the granularity of functions or system 
calls, a limitation that ALEA overcomes. Fine-grained energy profiling enables more
compile and run time opportunities for power-aware code optimization.

PowerPack~\cite{PowerPack} uses manual code instrumentation and platform-specific 
hardware instrumentation for component-level power measurement to associate power 
samples with functions.
NITOS~\cite{Keranidis:2014:NMM:2602044.2602047} measures energy consumption of 
mobile device components with a custom instrumentation device.
Similarly, LEAP~\cite{McIntire:2007:ESN:1236360.1236448} measures energy consumption 
of code running on networked sensors with custom instrumentation hardware.
These tools profile power at the hardware component level, thus capturing 
the power implications of non-CPU components, such as memories, interconnects, storage 
and networking devices. ALEA is complementary to these efforts. ALEA's sampling 
method can account for energy consumed by any hardware component between basic blocks,
while the statistical approach followed in ALEA overcomes the limitations of coarse
and variable power sampling frequency in system components.

Other energy profiling tools build instruction-level power models bottom-up 
from gate-level models, or other hardware models extracted at design time to
provide power profiles to simulators and prototyping 
environments~\cite{Tsoi:2011:PPO:2082156.2082159,Tu:2014:PPP:2597648.2566660}. 
These inherently static models fail to capture the variability in instruction-level 
power consumption due to the context in which instructions execute in real programs. 
Similarly, using microbenchmarks~\cite{Shao:2013:ECI:2648668.2648758} to estimate 
the energy per instruction (EPI) or per code block based  on its instruction mix 
does not capture the impact of the execution context.

\section{Platforms and Energy Measurement}
\label{sec:platforms}

The ALEA energy profiler builds on platform-specific substrates to measure or to model
power at a fine granularity based on data constrained by the sampling rate of the 
underlying power sensors. In this paper we use two distinct platforms for 
power measurement, one based on Intel's Running Average Power Limit (RAPL) apparatus 
on a Xeon Sandy Bridge  server and a second based on integrated power sensors on an 
ARM Exynos board.

On the Sandy Bridge server, we directly measure energy consumption through on-chip 
energy counters, which we access through the RAPL interface~\cite{intel_manual_rapl}.

RAPL allows us to account for the energy consumption of four components:
\texttt{PKG}, which measures the energy consumed by the processor package, 
including the multicore processor; \texttt{PP0}, which measures the energy consumed
by the power plane that powers the cores and the on-chip caches (L1/L2/L3); \texttt{PP1}, 
which measures the energy consumed by the on-chip graphics processor (for client 
platforms); and \texttt{DRAM}, which measures the energy consumed by memory DIMMs.

Client platforms can only access the \texttt{PKG}, \texttt{PP0} and \texttt{PP1} 
counters, while server platforms can access the \texttt{PKG}, \texttt{PP0} and 
\texttt{DRAM} counters. Our Sandy Bridge server includes two Intel Xeon E5-2650 
processors with eight cores per processor, 32KB/32KB I/D-Cache per core, 2MB shared 
L2 cache per 8 cores, and 20MB shared L3 cache per package. The system runs CentOS 
(release 6.5). The frequency of the system is up to 2 GHz. 
We disable the processor's Turbo Boost and Hyperthreading options in our validation experiments.

Our second platform, an ODROID-XU+E board, has one Exynos 5 Octa processor. 
This ARM Big.LITTLE architecture has four Cortex-A15 cores and four Cortex-A7 
cores, 32KB/32KB I/D-Cache per core, NEONv2 floating point support per
core, VFPv4 support per core, one PowerVR SGX 544 MP3 GPU, and 2 GBytes of LPDDR3 
DRAM. A 2 MByte L2 cache is shared between all Cortex-A15 cores and a 512 KByte L2 
cache is shared between all Cortex-A7 cores. The ODROID board also includes power 

meters on each voltage plane to measure consumption for the following four sets of 
components: Cortex-A7 cores, including their shared L2 cache; Cortex-A15 cores, 
including their shared L2 cache; GPU; and DRAM. The system runs Ubuntu 14.04 LTS. 
In our experiments, we use the Cortex-A15 cores only at their maximum frequency 
of 1.6 GHz.

\section{Profiling}
\label{sec:profiling}

Execution time profiling can use sampling or instrumentation~\cite{hpdctool_unwind,gwp}.
Compiler or binary instrumentation inserts profiling instructions that track dynamic 
execution counts and the execution time of code paths, as well as software or 
hardware events. Profilers based on sampling suspend binary execution to 
sample the execution state, typically the current program counter and possibly register 
contents or a stack traceback, and to correlate the sample with software events, 
hardware events, or metrics. 

\begin{figure}[t]
\centering
\includegraphics[width=\columnwidth, keepaspectratio]{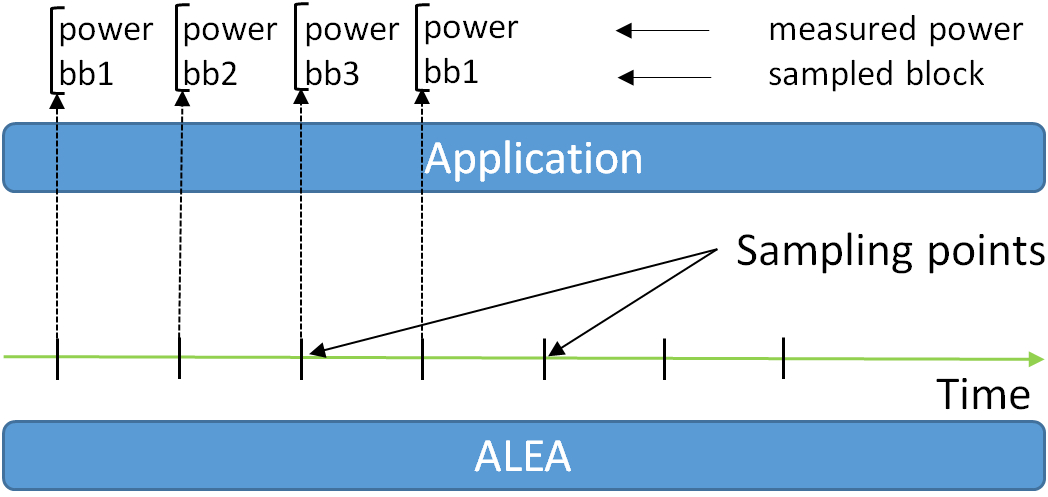} 
\caption{Sampling process}
\label{sampling_scheme}
\end{figure} 

We use statistical sampling for fine-grained energy profiling and demonstrate 
that we can probabilistically estimate energy consumption at fine and coarse 
granularities. Our profiling approach simultaneously samples the currently executing
basic block and takes power measurements, which it assigns to the basic block 
(Figure~\ref{sampling_scheme}). We perform a one-pass sampling of power measurements 
during a single program execution. Our tool processes the profiling results
off-line, using a probabilistic model to estimate the execution times and 
the mean power consumption for each basic block.  

\subsection{Execution time profiling model}
\begin{figure}[t]
\centering
\includegraphics[width=\columnwidth, keepaspectratio]{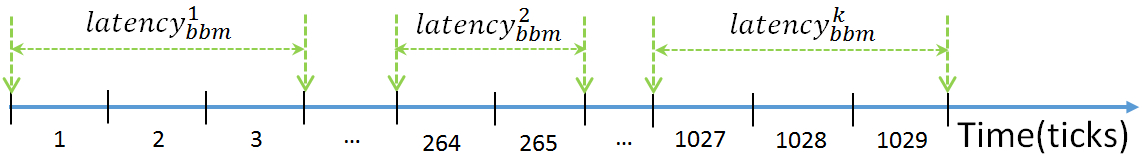} 
\caption{Execution of a basic block in a program}
\label{model_execution}
\end{figure} 

\begin{figure}[t]
\centering
\includegraphics[width=\columnwidth, keepaspectratio]{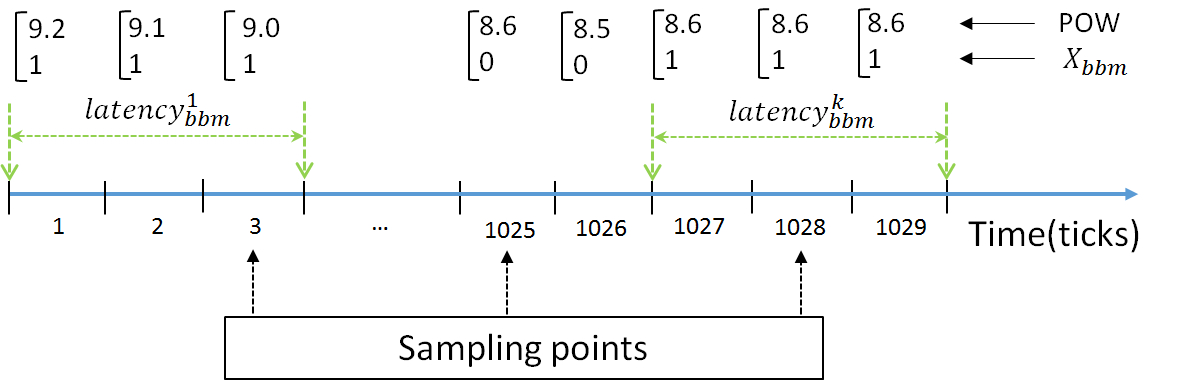} 
\caption{Random sampling}
\label{random_sampling}
\end{figure} 

To motivate the model, Figure \ref{model_execution} shows the iterative execution 
of a basic block that is executed $k$ times. The model makes the simplifying 
assumption that the processor executes instructions from one basic block ($bbm$) 
in each clock cycle. The latency of each basic block ($latency^{j}_{bbm}$) may vary 
between iterations. For example, a basic block may execute the same load instruction 
with different latencies between iterations, depending on the level of the memory 
hierarchy that provides the requested data. 

If we sample the program counter once during program execution at a random 
point in time,  we define the random variable $X_{bbm}$ as:
\begin{equation}
X_{bbm} = \begin{cases} 1, & \mbox{if } bbm \mbox{ is the sampled basic block} \\ 0, & \mbox{otherwise} \end{cases}
\end{equation}
In our probabilistic model, CPU clock cycles (ticks) correspond to the units of the 
finite population ($U$) and a sample during a specific clock cycle instantiates 
$X_{bbm}$~\cite{lohr:1999}. The probability that $bbm$ is sampled is:
\begin{eqnarray}
\label{eq:prob}
p_{bbm} = P(X_{bbm} = 1) = \frac{C^{1}_{t_{bbm}}}{C^{1}_{t_{exec}}} = \frac{\sum_{j=1}^k latency^{j}_{bbm}}{t_{exec}}\\
\frac{\sum_{j=1}^k latency^{j}_{bbm}}{t_{exec}} = \frac{t_{bbm}}{t_{exec}}
\end{eqnarray}

\noindent where $t_{bbm}$ is the total execution time of instances of $bbm$, $t_{exec}$ is 
the total execution time of the program, and $C^{1}_{S}$ is a 1-combination of a 
set $S$. We measure time in ticks
and represent it in seconds by dividing it by the CPU frequency. 
Equation~\ref{eq:prob} captures the observation that the probability of sampling
a basic block at a random clock cycle is equal to the ratio of its execution time 
to the program's total execution time. If the probability $p_{bbm}$ and the total execution 
time are known then $t_{bbm}$ is:
\begin{eqnarray}
t_{bbm} = p_{bbm} \cdot t_{exec}
\end{eqnarray}
We assume that $X_{bbm}$ follows a Bernoulli distribution because it is binary, 
random, and $p_{bbm}$ is a constant in our model. By applying random sampling (see Figure \ref{random_sampling}), we can estimate the probability 
as the maximum likelihood estimator of parameter $p_{bbm}$ in the Bernoulli distribution
for $X_{bbm} = 1$ \cite{CasBer90, montgomery2002applied}:
\begin{equation}
\hat{p}_{bbm} = \frac{n_{bbm}}{n}
\label{instr_prob_sample}
\end{equation}
In Equation~\ref{instr_prob_sample}, $n_{bbm}$ is the number of samples of some 
instruction from $bbm$, and $n$ is the total number of samples.
Thus, we estimate the execution time of any basic block as:
\begin{equation}
\hat{t}_{bbm} =  \hat{p}_{bbm} \cdot t_{exec} = \frac{n_{bbm} \cdot t_{exec}}{n}
\label{time_instr_estimate}
\end{equation}
We measure the total execution time $t_{exec}$ of an application during the profiling run.
\subsection{Energy profiling model}

We apply the same probabilistic approach to profile power and energy.
Similarly to the execution time profiling model, we consider power consumption 
as a random variable ($pow$, Figure~\ref{random_sampling}) and an implementation 
of this variable at a clock cycle as a characteristic associated with the clock 
cycle. We simultaneously take samples of the program counter and power consumption, 
which we assign to the sampled basic block even though power consumption likely 
includes power that instructions outside that basic block consume.

Assuming $n_{bbm}$ samples of block $bbm$, we estimate its mean power consumption as \cite{montgomery2002applied}:
\begin{eqnarray}
\label{eq:bbpower}
\hat{pow}_{bbm} = \frac{1}{n_{bbm}} \cdot \sum_{i=1}^{n_{bbm}} pow_{bbm}^{i}
\end{eqnarray}
In Equation~\ref{eq:bbpower}, $pow_{bbm}^{i}$ is the power consumption associated
with the $i-th$ sample of block $bbm$. 

We estimate the energy consumption of $bbm$ as: 
\begin{eqnarray}
\hat{e}_{bbm}  = \hat{pow}_{bbm} \cdot \hat{t}_{bbm}
\end{eqnarray}

\subsection{Bounds and Confidence}
If $p_{bbm}$ is not too close to $0$ or 
$1$ and $n$ is relatively large ($n \cdot p_{bbm} > 5$, $n \cdot (1-p_{bbm})> 5$)~\cite{montgomery2002applied}, 
then we can construct the confidence interval with upper and lower bounds on $p_{bbm}$:
\begin{equation}
\hat{p}^{u}_{bbm} = \hat{p}_{bbm} + z_{\alpha/2}\sqrt{\frac{1}{n} \cdot \hat{p}_{bbm} \cdot (1-\hat{p}_{bbm})}
\label{std_interval1}
\end{equation}
\begin{equation}
\hat{p}^{l}_{bbm} = \hat{p}_{bbm} - z_{\alpha/2}\sqrt{\frac{1}{n} \cdot \hat{p}_{bbm} \cdot (1-\hat{p}_{bbm})}
\label{std_interval2}
\end{equation}
\begin{equation}
\hat{p}^{l}_{bbm} \le p_{bbm} \le  \hat{p}_{bbm}^{u}
\label{std_interval3}
\end{equation}
In Equations~\ref{std_interval1},~\ref{std_interval2}~and~\ref{std_interval3}, 
$z_{\alpha}$ is the $1-\alpha/2$ percentile of the standard normal 
distribution, and $1-\alpha$ is a confidence level. The interval in Equation~\ref{std_interval3} 
includes the true value of $p_{bbm}$ with probability $1-\alpha$. According to 
Equation~\ref{time_instr_estimate}, by multiplying the lower and upper bounds of 
$p_{bbm}$ with the total execution time $t_{exec}$, we obtain an interval in which the 
true execution time $t_{bbm}$ of $bbm$ lies:
\begin{equation}
\hat{p}^{l}_{bbm} \cdot t_{exec} \le t_{bbm} \le \hat{p}^{u}_{bbm} \cdot t_{exec}
\end{equation}
We can similarly build a confidence interval for power~\citep{montgomery2002applied}:
\begin{equation}
\hat{pow}_{bbm}^{u} = \hat{pow}_{bbm} + z_{\alpha/2}\frac{s}{\sqrt{n_{bbm}}}
\label{std_power1}
\end{equation}
\begin{equation}
\hat{pow}_{bbm}^{l} = \hat{pow}_{bbm} - z_{\alpha/2}\frac{s}{\sqrt{n_{bbm}}}
\label{std_power2}
\end{equation}
\begin{equation}
 s = \sqrt{\frac{1}{n_{bbm}-1} \cdot \sum_{i=1}^{n_{bbm}} (pow_{bbm}^{i} - \hat{pow}_{bbm})^{2}}
\label{std_power3}
\end{equation}
\begin{equation}
\hat{pow}_{bbm}^{l} \le pow_{bbm} \le \hat{pow}_{bbm}^{u}
\label{std_power4}
\end{equation}
where $s$ is the corrected sample standard deviation. 
Using  confidence intervals for execution time and power, we can derive a confidence 
interval for energy consumption:
\begin{equation}
\hat{p}^{l}_{bbm} \cdot t_{exec} \cdot \hat{pow}_{bbm}^{l} \le e_{bbm} \le \hat{p}^{u}_{bbm} \cdot t_{exec} \cdot \hat{pow}_{bbm}^{u}	 
\label{std_energy}
\end{equation}
 
If we increase the total number of samples, we reduce the width of the confidence 
intervals as they are inversely proportional to the square root of the number of 
samples (time: $\sim\frac{const}{\sqrt{n}}$, power: $\sim\frac{const}{\sqrt{n_{bbm}}}$). Thus, the accuracy of the energy 
estimates should increase with increasing total number of samples ($n$) and the 
given basic block samples ($n_{bbm}$). Because $n_{bbm}$ is strongly correlated 
with $n$, the accuracy of the energy estimates is primarily affected by the 
total number of samples ($n$).   

\subsection{Profiling of parallel applications}
We employ the same execution time and energy profiling models for multithreaded 
applications. The essential difference is that each sample is a vector 
of program counters simultaneously sampled across all threads.
Thus, we distribute the execution time and energy 
across combinations of basic blocks, which are executed on different threads:
\begin{eqnarray}
\hat{t}_{comb} = \hat{p}_{comb} \cdot t_{exec} = \frac{n_{comb} \cdot t_{exec}}{n}
\label{time_instr_estimate_par}
\end{eqnarray}
\begin{eqnarray}
\hat{pow}_{comb} = \frac{1}{n_{comb}} \cdot \sum_{i=1}^{n_{comb}} pow_{comb}^{i}
\label{power_instr_estimate_par1}
\end{eqnarray}
\begin{eqnarray}
comb = bb_{thread_{1}},bb_{thread_{2}},...,bb_{thread_{l}}
\label{power_instr_estimate_par2}
\end{eqnarray}
where $comb$ corresponds to a combination of basic blocks that were sampled 
on different threads ($l$ threads).

We consider all threads of an application running on the same processor package
collectively during sampling, because they share resources and because resource sharing contributes additional energy 
consumption due to contention between threads. Shared resources include caches, 
buses and network links, all of which can significantly increase power consumption 
under contention. We could apportion power between threads based on dynamic activity 
vectors that measure the occupancy of shared hardware resources per thread~\cite{Manousakis2014}.
However, these vectors are difficult to collect on real hardware, as current 
monitoring infrastructures cannot distinguish between the activity of different 
threads on shared resources. As such, per-thread energy apportioning cannot be 
accurately validated on real hardware.

We can still correlate power consumption with basic blocks with this approach. For 
example, we can investigate how the energy profile of a basic block changes between 
stand-alone execution and execution with different co-runners, to capture 
contention for shared resources. 
Further, our methodology helps us understand how synchronization can decrease power 
consumption, which in turn reveals opportunities for reducing energy consumption in 
the runtime system by applying dynamic concurrency throttling~\cite{Li:2013:SER:2420628.2420808}.

\subsection{Power measurements}

We measure processor power consumption on our Sandy Bridge server for a given sample 
($pow_{bbm}^{i}$) by dividing the energy consumed since the last sample by the length
of the sampling period. Our analysis of sampling overhead and accuracy, which we 
present in the following sections, led to a 10 ms sampling period. This approach 
conforms to RAPL, which provides running energy but not power measurements.

Our Exynos platform has \texttt{TI INA231} power meters, which 
directly sample power consumption for the system-on-chip averaged over a user-defined
period. We used the minimum feasible period on the Exynos, which is 280 microseconds.

In general the sampling period used in our model is different than the platform
power sampling period. 
Our  method estimates the energy consumption of basic blocks of any duration,
including ones that run for less than the sampling period, under a 
probabilistic model of the fraction of program execution time that each given basic 
block consumes and the average power consumption due to execution of that basic block.
\subsection{Implications of systematic sampling}
Systematic sampling, which approximates random sampling, selects units from an 
ordered population with the same sampling period. It selects the first unit of 
a sample randomly from the bounded 
interval $ [1,length\ of\ sampling\ period]$. We use systematic sampling for time 
and energy profiling, in which units correspond to CPU clock cycles and the user 
sets the sampling period~\cite{lohr:1999}.

Systematic sampling can be inefficient with populations that exhibit a 
periodic variation that is an integral multiple of the 
sampling period. For example, if the same basic block is executed with a period 
equal to the sampling period then theoretically, we will only sample that basic block.
In practice, the precise size of a sampling period in CPU clock cycles varies 
randomly between samples due to the inaccuracy of the timer and variance in the 
execution length of the sampling code itself. We find that on the Sandy Bridge 
and Exynos platforms, the variation in the delay between samples may be 
up to hundreds of microseconds. This random variation obviates the need to add 
deliberate randomization during the sampling process.

\subsection{Sampling period}

The accuracy of our sampling estimates improves with an increasing 
number of samples. However, sampling incurs overhead, which biases execution time 
and energy estimates. This overhead increases linearly or superlinearly with the 
number of samples, since the program must be interrupted for each sample. Thus, the 
estimation error is composed of random error, which is introduced by sampling, and 
systematic error, which is introduced by profiling overhead. If we increase the 
number of samples, then the random error decreases but the systematic error increases.

\begin{figure*}[t]
\centering
\includegraphics[keepaspectratio=true,width=\textwidth]{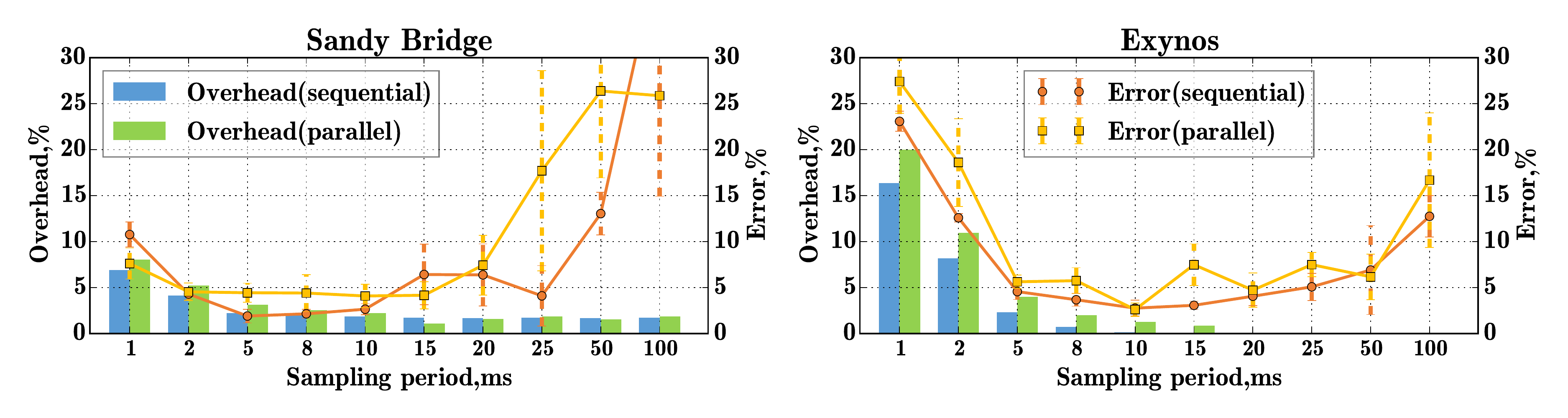}
\caption{Overhead and energy estimate error}
\label{figure_overhead_vs_error}
\end{figure*}

We use our benchmark suite to capture basic blocks with diverse execution 
times and power consumption to find the best sampling period in terms of energy 
estimation accuracy and execution time overhead. As an example, the 
\texttt{streamcluster} benchmark from the \texttt{Rodinia} suite includes basic 
blocks with latency varying between 1 and 30 ms on the Sandy Bridge platform.
Figure~\ref{figure_overhead_vs_error}
shows the trade off between the length of the sampling period, overhead and 
accuracy of energy estimates for the Sandy Bridge and Exynos platforms, using both
sequential and parallel executions of the benchmark. We observe similar results
in all benchmarks, pointing to a sampling period of 10 ms as a good compromise
between energy estimation error and runtime overhead. A fixed sampling 
period helps deployment of ALEA as a continuous, online application energy profiler 
with capped overhead. However, we can select an application-specific sampling 
frequency since the tool exposes the sampling interval as a user-defined parameter.

\subsection{Implementation}

ALEA uses a separate control process to obtain the current instruction pointer 
of the profiled application and to take power measurements. We use the \texttt{ptrace}
interface, which allows one process to retrieve the contents of registers in another 
process or thread. Thus, the profiled program does not execute any additional code, 
unlike sampling schemes based on signals~\cite{hpctool_analysis}. 
Instead, the control process captures context information and energy/power 
measurements. This approach reduces system overhead because system call interfaces 
are offloaded from the profiled program's critical path to the control process. 
However, this approach still incurs performance and energy overhead because processes
or threads of the profiled program are suspended while the control process reads 
the registers via the \texttt{ptrace} interface. 
 
ALEA currently executes on a dedicated core that the profiled application does not
use.

\section{Validation}
\label{sec:validation}

\begin{figure*}[t]
\centering
\includegraphics[keepaspectratio=true,width=\textwidth]{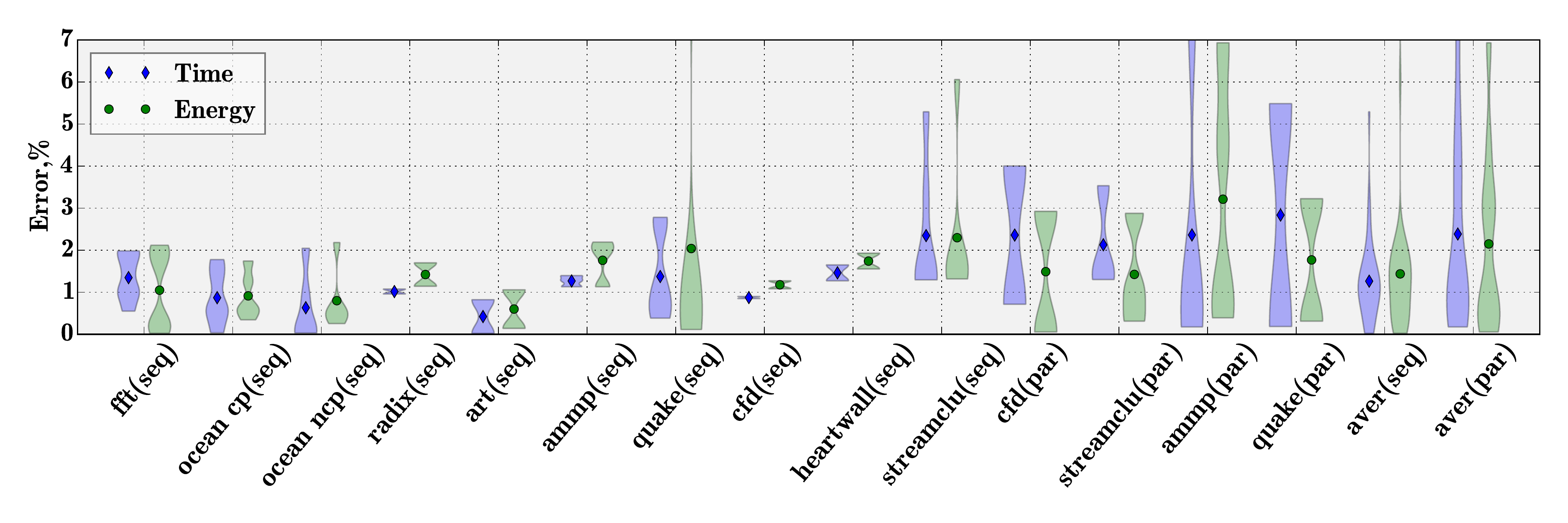}
\caption{Average error in execution time and energy estimates, compared with direct measurements (Sandy Bridge)}
\label{figure_validation_general}
\end{figure*}

We use 14 benchmarks (sequential and parallel) from four suites 
(\texttt{SPEC 2000}, \texttt{Parsec}, \texttt{Rodinia}, \texttt{SPEC OMP}) to 
validate the accuracy of ALEA's execution time and energy consumption estimates. 
We use a range of benchmarks to achieve good coverage of basic block features such
as execution time, including fine-grain and coarse-grain blocks, and energy 
consumption, including blocks with distinct power profiles and/or power variations 
between their samples. We use the \texttt{native} input data set for benchmarks 
from \texttt{Parsec} and  standard input for benchmarks from other suites.

We measure whole program execution time and energy. We also measure the execution 
time and energy of those basic blocks with latency that exceeds the sampling period 
(10 ms) in isolation. Further, in isolation, we measure the execution time and energy
of fine-grain basic blocks that have shorter latency than the sampling period,
but are enclosed in innermost loops such that the overall loop latency exceeds 
this period. Overall, direct per-basic block measurements covers $81\%$ of the 
execution time of each benchmark on average. We compare ALEA's execution time and 
energy consumption estimates to per-basic block direct measurements. For basic 
blocks that are not captured by direct measurements, we compare whole program 
measurements to the sum of execution time and energy consumption estimates for all 
basic blocks sampled by ALEA at least once during program execution.

We execute each benchmark at least six times. The first run directly measures 
energy and time. The other runs use ALEA to estimate the execution time and 
energy consumption of each basic block.  We use at least five ALEA runs and 
as many more as needed (up to 20 total) to bring the 95\% confidence interval 
of the time, power and energy measurements within 5\% of the mean.
We compile all benchmarks using \texttt{gcc} with \texttt{-O1} 
and \texttt{-ffast-math}, which inlines mathematical and other functions when 
possible. For validation, we use the \texttt{-O1} optimization level instead of 
\texttt{-O3} to increase latencies of some basic blocks to the minimum needed 
to take direct measurements. 

The ALEA profiler executes on a core that is not in use by the 
profiled application, to minimize interference. Specifically, ALEA runs on a separate Sandy Bridge
socket but on the same Exynos four-core Cortex A15 cluster since our Odroid board 
does not allow co-execution on both of the A15 and A7 clusters.  
We present results from experiments using up
to eight threads on one socket of the Sandy Bridge platform and up to two threads of
the A15 cluster on the Exynos platform for the execution of parallel benchmarks. Running the profiler
on a separate core keeps the overhead under 1\% on both platforms. We also experimented with
running the profiler on the same core as one of the threads of each profiled program and observed 
the overhead to increase to up to 10\% (not shown). This overhead can be mitigated by reducing the sampling
frequency (Figure~\ref{figure_overhead_vs_error}).
Halving the sampling frequency halves the overhead and keeps the ALEA average energy estimation error 
at a manageable 5\% (Exynos) to 6\% (Sandy Bridge).

\subsection{Sandy Bridge results}

Figure~\ref{figure_validation_general} presents the average error of ALEA's execution
time and energy consumption estimates for basic blocks on the Sandy 
Bridge platform. The average error is 1.3\% for the execution time estimates and 
1.4\% for the energy consumption estimates. 99\% of the execution 
time and energy measurements lie within 95\% confidence intervals. 
For those fine-grain basic block sets enclosed in loops that allow us to measure time 
and energy directly, the average error in ALEA's energy estimate is 1.6\% (1.3\% 
for execution time). For coarse-grain basic blocks, the ALEA profiling error is 
1.4\% for both execution time and energy consumption. 
The average errors of the ALEA execution time and energy estimates for parallel 
benchmarks (Figure~\ref{figure_validation_general}) are 3.1\% and 2.6\%. 
Our average whole program absolute error across all benchmarks is 1.1\% for
execution time and 1.4\% for energy.

\subsection{Exynos results}

While RAPL supports direct energy measurements on the Sandy Bridge platform,
we can only directly measure power on the Exynos platform. We thus follow a different
approach to validate energy profiling between basic blocks on it. We again instrument
the benchmarks to perform execution time profiling. However, in each instrumented 
basic block, we sample the power consumption using the system timer and corresponding
signal handler. We set the Exynos TI power meters to compute average power over the 
minimum feasible period of 280 microseconds. This instrumentation has higher overhead than direct 
energy measurements on the Sandy Bridge platform because it enforces one interrupt per sample. 
This higher overhead introduces a bias in energy measurements, which leads to higher 
error.

The average error in ALEA's energy estimates (not shown due to space limitations) 
is 2.6\% (also 2.6\% in execution time estimates) for sequential 
benchmarks and 3.6\% (2.8\% in execution time estimates) for parallel benchmarks. 99\% of all 
time and energy measurements lie within 95\% confidence intervals. The average 
error in ALEA's energy estimate for fine-grain basic blocks is 3.5\% (3.7\% for 
execution time) and 1.9\% (1.8\% for execution time) for coarse-grain basic blocks.
The average error of total execution time estimates is 1.4\% and that of total energy estimates is 1.9\%.

\section{Impact of Memory Instructions and Synchronization on Energy}
\label{sec:experiments}
\begin{figure}[t]
\centering
\includegraphics[keepaspectratio=true,width=\columnwidth]{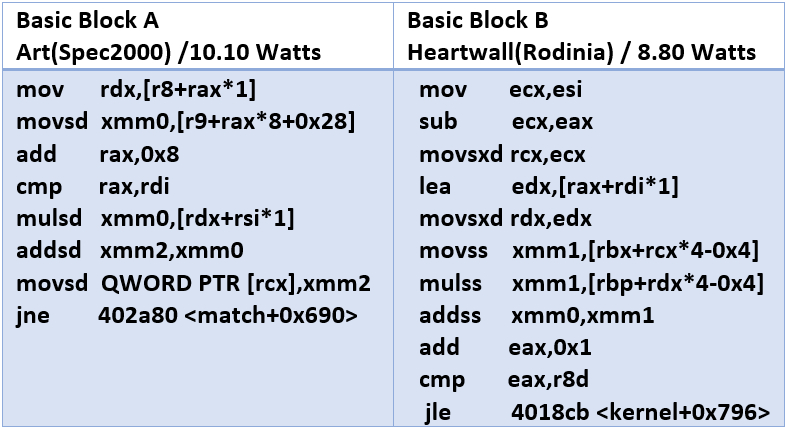} 
\caption{\texttt{art} and \texttt{heartwall} basic blocks (Sandy Bridge)}
\label{art_vs_heartwall_bb}
\end{figure}
\begin{figure*}[t]
\centering
\includegraphics[keepaspectratio=true,width=\textwidth]{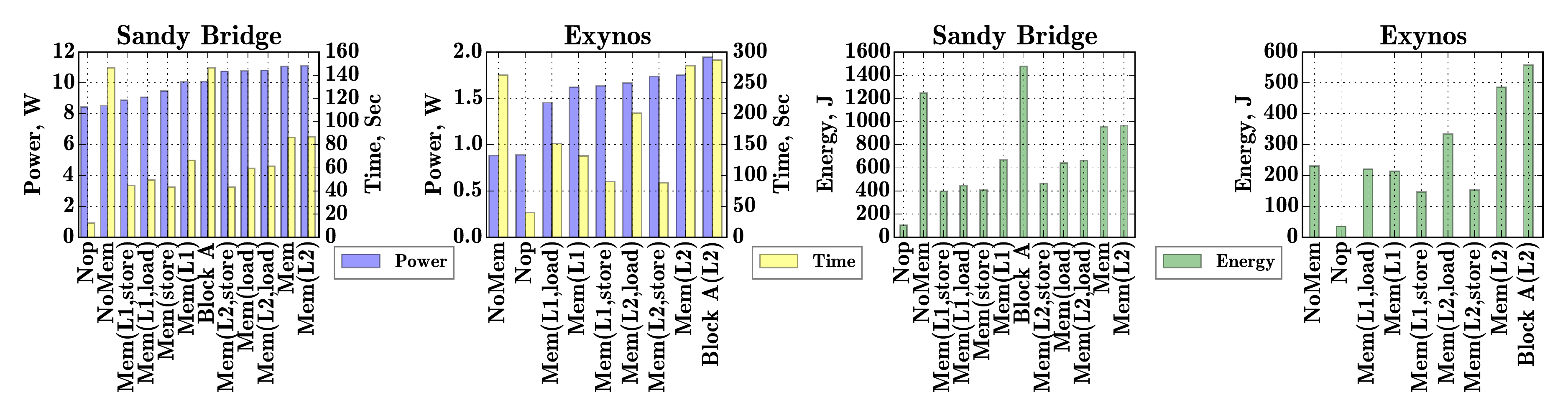}
\caption{Power,energy and execution time measurements taken for microbenchmarks}
\label{figure_power_basicA}
\end{figure*}
We can optimize a program's energy consumption by reducing its execution time or 
power consumption. However, reducing execution time often increases power consumption.
We use ALEA to investigate the causes of increased power consumption 
in optimized programs. Our experiments indicate that the power consumption may vary 
considerably between basic blocks. Figure~\ref{art_vs_heartwall_bb} shows a basic 
block from \texttt{art} (\texttt{BBA}) and a basic block from \texttt{heartwall} 
(\texttt{BBB}). On the Sandy Bridge platform \texttt{BBA} consumes 10.10W
(98.39J in total), while \texttt{BBB} consumes 8.80W 
(278.63J in total). Our experimental study shows that the power consumption 
of a basic block is primarily affected by the cache access intensity and does not 
vary considerably with the type of executed instructions. In our example, \texttt{BBA}
accesses approximately 7 MB of data during its execution (which fits in the L3 cache),
while \texttt{BBB} accesses only 36KB of data (which fits in the L1 cache).
The Exynos platform exhibits similar behavior.

\begin{table}
\footnotesize
\begin{center}
\begin{tabular}{ |c|c| } 
 \hline
 \textbf{Block} & \textbf{Description}\\ 
 \hline
 \texttt{Basic block} A & Copy of \texttt{BBA}\\ 
 \hline
 \texttt{Mem} & Only memory access instructions of \texttt{BBA} \\
 \hline
 \texttt{NoMem} & Only arithmetic/logic instructions of \texttt{BBA} \\
 \hline
 \texttt{Mem(L2)}& \texttt{Mem} block with the size of accessed \\
         & data limited to 2MB (L2 cache size on Exynos)\\
 \hline
 \texttt{Mem(L1)}& \texttt{Mem} block with the size of accessed \\
         & data limited to 2KB (L1 cache size on Exynos)\\
 \hline
 \texttt{Mem(load)} & \texttt{Mem} block with load instructions only \\ 
 \hline
 \texttt{Mem(store)} & \texttt{Mem} block with store instructions only\\ 
 \hline
 \texttt{Mem(L2,load)}& \texttt{Mem(L2)} block with loads only\\
 \hline
 \texttt{Mem(L2,store)}& \texttt{Mem(L2)} block with stores only\\
 \hline
 \texttt{Mem(L1,load)}& \texttt{Mem(L1)} block with loads only\\
 \hline
 \texttt{Mem(L1,store)}& \texttt{Mem(L1)} block with stores only\\
 \hline
\end{tabular}
\end{center}
\caption{Versions of \texttt{BBA}
\label{versions_ob_basic_blockA}}
\end{table}

To confirm the effect of cache accesses, we develop microbenchmarks based on 
\texttt{BBA}. We create a basic block with the same set of instructions 
and context for both processors. We divide its instructions into two groups: 
memory access instructions and arithmetic/logic instructions. We use these groups 
to implement different versions of \texttt{BBA} (Table~\ref{versions_ob_basic_blockA}).
We then add a basic block with a single \texttt{nop} instruction, which does not 
use the floating point units (FPUs). We limit the size 
of the accessed data so that the data fits in the L2 cache.

Figure \ref{figure_power_basicA} shows the power, execution time and energy measurements for 
our experimental set of basic blocks on the Sandy Bridge platform (the basic blocks are sorted by power consumption). The \texttt{Nop} 
and \texttt{NoMem} blocks consume almost the same power even though the second block 
occupies the FPU. In contrast, the difference in power consumption between the 
\texttt{Mem} and \texttt{NoMem} blocks is more than 1.5W. Similarly to the Sandy 
Bridge platform, the \texttt{Nop} and \texttt{NoMem} basic blocks show the same power
consumption on the Exynos platform, while the \texttt{Mem (L2)} block consumes more 
power than does the \texttt{NoMem} block (Figure~\ref{figure_power_basicA}). Thus,
the increase in power consumption on both platforms is primarily due to data cache 
accesses and not the type of instructions executed.Even though the \texttt{NoMem} block merely omits the memory access instructions
of \texttt{BBA}, these blocks have nearly the same execution time on both platforms 
because pipelining hides the data access latencies of \texttt{BBA}. Thus, its 
execution time does not increase despite the energy used for the data accesses.

Pipelining can lead to significant errors in energy consumption estimates based on 
EPI~\cite{Shao:2013:ECI:2648668.2648758}, which ALEA mitigates. For example, \texttt{BBA} is a union of 
instructions from \texttt{Mem} and \texttt{NoMem} blocks. On the Sandy Bridge, 
according to an EPI model, \texttt{BBA}, which consumes 1,474J, 
should consume the sum of the energy consumed by \texttt{Mem} (955J) and 
\texttt{NoMem} (1,245J) blocks, which is  2,200J or over 1.5$\times$ more than the actual
energy consumption. On the 
Exynos platform, the energy consumption of \texttt{BBA} is 1.29$\times$ less 
than the sum of energy consumption of the \texttt{NoMem} and \texttt{Mem} blocks.

Our experiments show that the power consumption of basic blocks executed in parallel 
applications depends on the form of each thread's activity. For example, the 
\texttt{ammp (SPEC OMP)} benchmark contains a basic block with 564 instructions that
correspond to a loop body in the \texttt{ mm\_fv\_update\_nonbon} procedure 
(\texttt{rectmm.c}, line 1210). This basic block includes regular accesses to caches.
When four threads execute this block in parallel,  the Sandy Bridge processor 
consumes \texttt{19.07W} (\texttt{1153J}). 
However, if only one thread executes this basic block while the other threads 
wait in synchronization, power consumption drops to \texttt{13.19W} 
(\texttt{513J}). Results on the Exynos platform are similar.

\section{Use cases}
\label{sec:use_case}

\begin{figure*}[t]
\centering
\includegraphics[keepaspectratio=true,width=\textwidth]{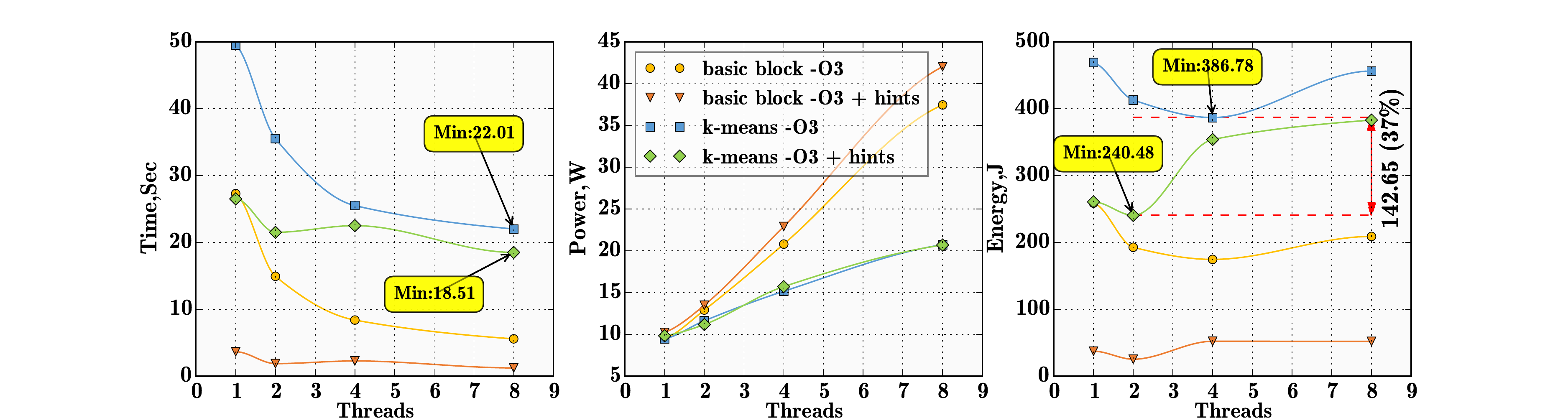}
\includegraphics[keepaspectratio=true,width=\textwidth]{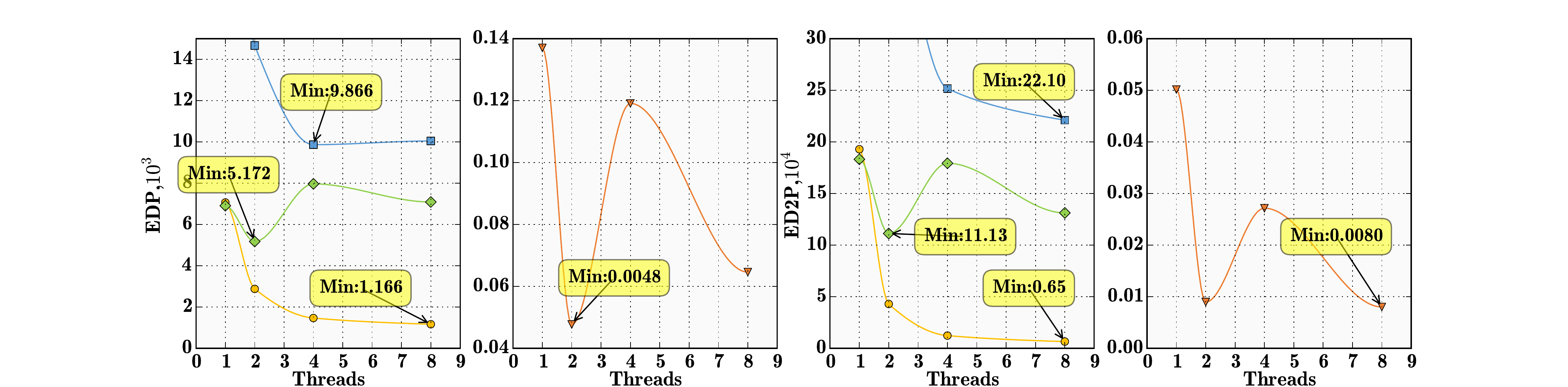}
\caption{Profiling results of k-means (Sandy Bridge)}
\label{figure_results_kmeans}
\end{figure*}

We present three use cases of how basic block level energy profiling can be used 
in energy-aware program optimization. Our first use case analyzes hot spots to 
uncover opportunities for energy optimizations in a single dominant basic block, 
based on techniques that adapt the degree of parallelism in the program~\cite{Curtis-Maury:2006:OPA:1183401.1183426,Curtis-Maury:2008:PPA:1449379.1449456,Jeon:2013:APW:2465351.2465367,Sridharan:2014:AEP:2594291.2594292}.
Our second and third use cases explore fine-grain optimization and power capping opportunities 
across multiple basic blocks. 
\subsection{Hotspot energy optimization}

Our first use case applies ALEA to optimize hot spot energy use in the 
\texttt{k-means} benchmark of the \texttt{Rodinia} suite using one socket 
on our Sandy Bridge platform. ALEA runs on one core of the other 
socket. We scaled up the standard input set $6\times$ to model 
realistic runs of the benchmark. Profiling of the sequential version 
shows that 56\% of the total execution time is spent on the basic block that
corresponds to the loop that calculates the multidimensional spatial Euclidean 
distance square (\texttt{euclid\_dist\_2} function). We use the \texttt{-O3} 
compilation flag as a default option. However \texttt{unroll} 
and \texttt{auto-vectorization} optimizations are, surprisingly, not applied to 
the basic block. We use compiler hints (C-extensions: parameter and function 
attributes) to force the compiler to apply unrolling. We also use parameter 
attributes to align and to restrict pointers so the compiler recognizes the proper 
context for auto-vectorization. Finally, the \texttt{-ffast-math -mavx} flag 
enables floating-point arithmetic transformations and the use of \texttt{AVX-256} 
instructions. We refer to this set of optimizations as \textit{hints}.

Figure \ref{figure_results_kmeans} shows execution time, power, energy, energy-delay
and energy-delay$^{2}$ estimates for the key \texttt{k-means} basic block optimized 
with \texttt{-O3}  and with \texttt{-O3+hints}. The energy-delay and 
energy-delay$^{2}$ measurements of the latter version are shown in separate charts 
to assist the reader, because the optimization hints reduce these metrics 
by two to three orders of magnitude. We also measure the corresponding metrics for 
the entire  \texttt{k-means} program. Our optimizations reduce execution time of the 
dominant basic block by up to 8$\times$ when running with one or two threads but the 
impact of these optimizations on performance is less pronounced with more 
threads, due to memory contention that limits 
scalability. The speedup of the full benchmark when running with more cores is 
limited by the significant percentage of sequential execution time spent on I/O 
operations (up to 55\% after optimizations). The optimization hints that 
significantly accelerate the dominant basic block actually reduce the speedup 
from using more cores.

\begin{table*}
\small
\begin{center}
\begin{tabular}{ |c|c|c|c|c|c|c|c| }
 \hline
 & \multicolumn{2}{|c|}{\textbf{Baseline}} & \multicolumn{5}{|c|}{\textbf{Energy-optimal}}\\
 \hline
 & Time(s) & Energy (J) & Time (s) & Energy (J) & Threads & Frequency & Manual optimization \\
 \hline
bb1,jacobcalc2.C:301& 2.03 & 8.48 & 1.87 & 6.03 & 4 & 1500 MHz & No \\
bb2,slave2.C:641 & 1.54 & 6.70 & 1.31 & 4.16 & 2 & 1600 MHz & Yes \\
bb3,laplacalc.C:83 & 2.02 & 9.53 & 2.55 & 7.98 & 2 & 1500 MHz & No \\
bb4,multi.C:253 & 2.17 & 7.22 & 2.62 & 6.52 & 2 & 1500 MHz & No \\
bb5,multi.C:235 & 2.36 & 7.88 & 3.29 & 5.56 & 1 & 1500 MHZ & No \\
bb6,multi.C:290 & 2.67 & 9.23 & 3.23 & 5.46 & 1 & 1500 MHz & No \\
\hline
program & 29.93 & 108.64 & 26.88 & 72.84 & 2.0 (avg.) & 1516 MHz (avg.) & Yes \\
 \hline
 \end{tabular}
\end{center}
\caption{ Time and energy impact of basic-block level optimization for \texttt{ocean\_cp} on Exynos}
\label{oceancp_results}
\end{table*}

The impact of optimizations on energy consumption is considerably different from that on execution 
time. Power consumption increases disproportionally when optimizations and additional
concurrency are applied to the benchmark. Energy consumption is not minimized with 
the set of optimizations or the degree of concurrency that minimizes execution time. 
A combination of unrolling, vectorization and maximal concurrency (eight threads) 
achieves peak performance for the benchmark (18.51 seconds), while energy consumption 
is minimized with optimizations turned on but using only two cores, at a 
20\% performance loss. Overall, optimizing the dominant basic block for energy
consumption yields 37\% energy savings for the entire program, compared to the
high-performance baseline (eight cores, -O3 + hints).

The \texttt{k-means} example exhibits clear trade-offs between performance and energy
consumption. Optimization criteria that place heavier emphasis on performance (execution
time, energy-delay$^{2}$), when applied to the dominant basic block, indicate
preference for the highest concurrency and manual code optimization via hints.
Optimization criteria that place heavier emphasis on power and energy opt for
lower concurrency. Further, we should apply a different optimization strategy 
for the whole of the program, compared to the strategy followed for the dominant basic block
(see $EDP$ and $ED2P$ in Figure \ref{figure_results_kmeans}, configurations are annotated). 
This result motivates fine-grain energy accounting.

\subsection{Fine-grain power optimization across basic blocks}

We use the \texttt{ocean\_cp} benchmark from the \texttt{PARSEC} suite to explore 
whether ALEA exposes different energy optimizations for basic 
blocks in the same code, in order to achieve better whole-program energy-efficiency. 
Such an optimization strategy would motivate ALEA's fine-grain profiling.  We use 
the \texttt{native} input data set and modify the time between relaxations to 
increase the overall execution time of the benchmark in order to achieve stable
and repeatable results.  Time profiling of \texttt{ocean\_cp} indicates that more 
than 50\% of the total execution time is spent executing six basic blocks 
(Table~\ref{oceancp_results}), to which we refer as  \texttt{bb1} through 
\texttt{bb6}. We initially compile this benchmark for highest performance using 
the flags:\texttt{-O3}, \texttt{-mfpu=neon-vfpv4}, 
\texttt{-mtune=cortex-a15}, \texttt{-ffast} \texttt{-math}, 
\texttt{-funroll}\texttt{-loops}, \texttt{-ftree} \texttt{-vectorize},\\ 
\texttt{-fprefetch} \texttt{-loop}\texttt{-arrays}.

Motivated by our experimental analysis of the power implications of memory 
instructions (Section~\ref{sec:experiments}), we disable optimizations that 
could increase cache access rates to reduce power. The disabled 
optimizations are prefetching, for \texttt{bb3}, and the combination of 
unroll and vectorization, for \texttt{bb1} and \texttt{bb2}. By disabling 
these optimizations for those basic blocks, we reduce power consumption
by up to 14\% for \texttt{bb2}, 10\% for \texttt{bb1}, and 4\% for \texttt{bb3}.
Further code inspection of \texttt{bb4}, \texttt{bb5} and \texttt{bb6} reveals 
that the compiler inserts additional stack access instructions before each of 
these basic blocks, due to the predictive commoning optimization, which has no 
effect on performance, but increases power consumption. By disabling this 
optimization we reduce power consumption for these three basic blocks by 
between 3\% to 10\%.

Table~\ref{oceancp_results} shows selected results from an experimental campaign 
to understand how to minimize the energy consumption of the six dominant basic 
blocks in \texttt{ocean\_cp}. The baseline for this campaign is execution of the 
code using the maximum number of  cores on an Exynos cluster (four) and the maximum 
frequency (1600 MHz). Besides execution time and energy of the baseline case, we 
show execution time and energy of the energy-optimal configuration, as well as 
details of the program and system configurations that achieve energy minimization, 
including clock frequency, number of threads and use or no use of the three manual 
power optimizations considered: unrolling, vectorization and predictive commoning. 

The table reveals several findings that motivate the ALEA approach to fine-grain 
profiling. First, fine-grain energy optimization at the basic block level yields 
substantial energy savings, ranging from 10\% for \texttt{bb4} to 41\% for 
\texttt{bb6}; and 33\% for the program as a whole compared to the baseline. Second, the factor that 
catalyzes energy minimization varies between basic blocks: most basic blocks 
are more energy-efficient when running at slightly lower than the maximum 
frequency (1500 vs.\ 1600 MHz); most basic blocks run most efficiently with 
one or two, not all four, cores on the chip, suggesting that system 
bottlenecks such as memory contention dominate energy consumption; and at least 
one basic block (\texttt{bb2}) requires manual optimization to achieve maximum 
energy-efficiency. Third, fine-grain power optimization implies the ability to 
perform fine-grain power capping and more efficient power-constrained execution 
beyond that afforded by voltage and frequency scaling.
For example a 10\% reduction of the power cap in Exynos can be met by reducing 
frequency by one step but also by concurrency throttling and manual or 
compiler-driven code optimization. The latter two options show better 
energy savings potential. 

\subsection{Optimization of fine-grain basic blocks in acyclic regions}
Loops enclose all basic blocks considered in our other use cases. However, 
applications, such as the \texttt{Raytrace} benchmark from the PARSEC suite,
often contain hot basic blocks in acyclic regions. With the  \texttt{simlarge} 
input, the \texttt{SphPeIntersect} function, which contains two hot blocks in 
an acyclic region (lines 323--328, lines 333-335, sph.C) consumes about 50\% of 
the total execution time on the Exynos platform. The compiler optimizes these blocks poorly, leading to 
redundant memory accesses and indirect addressing instructions. We manually 
modified the generated code to remove redundant instructions, which reduced 
total energy consumption of the sequential version by 6.1\% (2.8\% for the 
parallel version). 

We cannot directly profile the targeted basic blocks due to the latency of 
hardware energy measurements. The execution time of the \texttt{SphPeIntersect} 
function is no more than 200 cycles on average. 
ALEA's probabilistic model was the only viable option to profile and to optimize 
these basic blocks.

\section{Conclusion}
\label{sec:conclusion}

We presented a probabilistic approach for fine-grained energy profiling,
implemented in ALEA, an energy profiling tool 
based on statistical sampling.
We demonstrated that fine-grain energy accounting provides better insight into the 
power implications of microarchitectural and memory structures to support 
energy-aware code optimization. ALEA importantly overcomes the fundamental
limitation of the low sampling frequency of power sensors, which is common across
computing platforms. The tool operates entirely in user space and is portable across 
architectures.

We demonstrated ALEA's high accuracy and low overhead on an Intel 
and an ARM platform with radically different architectural characteristics. ALEA achieved both functional and performance portability.

We used ALEA to demonstrate the strong correlation between power consumption and 
memory access rates, as well as a clear impact of shared cache contention on power consumption.  We presented use cases of 
ALEA where we applied new energy optimizations of individual basic blocks, using 
different strategies and achieved whole-program energy savings of
up to 37\%. These use cases motivated fine-grain energy accounting and 
uncovered the complex interplay between code optimization, multicore execution 
and energy consumption.

We will pursue three directions for future work in ALEA. The first direction is to
evolve ALEA into a production-strength energy accounting tool that maps energy 
consumption to source code and data structures, along the lines of tools such as Intel's 
Vtune and HPCToolkit. The second direction is to extend ALEA's 
capabilities to provide binary-level energy accounting of legacy programs 
running on virtualized software stacks.  The third direction is to use ALEA for  
constructing a new library of code optimizations for power-constrained environments.

\section*{Acknowledgment}
This research has been supported by the UK EPSRC through grant agreements EP/L000055/1 (ALEA), EP/L004232/1 (ENPOWER), and EP/K017594/1 (GEMSCLAIM) and by the EC 
FP7, through grant agreements FP7-610509 (NanoStreams) and FP7-323872 (SCORPIO).

\bibliographystyle{IEEEtran}
\bibliography{IEEEabrv,alea_pact}

\begin{thebibliography}{34}
\providecommand{\natexlab}[1]{#1}
\providecommand{\url}[1]{\texttt{#1}}
\expandafter\ifx\csname urlstyle\endcsname\relax
  \providecommand{\doi}[1]{doi: #1}\else
  \providecommand{\doi}{doi: \begingroup \urlstyle{rm}\Url}\fi

\bibitem[ina()]{ina231}
\emph{High- or Low-Side Measurement, Bidirectional CURRENT/POWER MONITOR with
  1.8-V I2CTM Interface}.

\bibitem[Adhianto et~al.(2010)Adhianto, Banerjee, Fagan, Krentel, Marin,
  Mellor-Crummey, and Tallent]{hpctoolkit}
L.~Adhianto, S.~Banerjee, M.~Fagan, M.~Krentel, G.~Marin, J.~Mellor-Crummey,
  and N.~R. Tallent.
\newblock {{HPCToolkit}: {Tools} for Performance Analysis of Optimized Parallel
  Programs}.
\newblock \emph{Concurr. Comput.: Pract. and Exper.}, 22\penalty0 (6):\penalty0
  685--701, Apr. 2010.
\newblock ISSN 1532-0626.
\newblock \doi{10.1002/cpe.v22:6}.
\newblock URL \url{http://dx.doi.org/10.1002/cpe.v22:6}.

\bibitem[Alonso et~al.(2013)Alonso, Dolz, Mayo, and
  Quintana-Ort\'{\i}]{Alonso:2013:EED:2558756.2558785}
P.~Alonso, M.~F. Dolz, R.~Mayo, and E.~S. Quintana-Ort\'{\i}.
\newblock {Energy-Efficient Execution of Dense Linear Algebra Algorithms on
  Multi-core Processors}.
\newblock \emph{Cluster Computing}, 16\penalty0 (3):\penalty0 497--509, Sept.
  2013.
\newblock ISSN 1386-7857.
\newblock \doi{10.1007/s10586-012-0215-x}.
\newblock URL \url{http://dx.doi.org/10.1007/s10586-012-0215-x}.

\bibitem[Bertran et~al.(2013)Bertran, Gonzalez~Tallada, Martorell, Navarro, and
  Ayguade]{Bertran:2013:SMG:2498752.2499068}
R.~Bertran, M.~Gonzalez~Tallada, X.~Martorell, N.~Navarro, and E.~Ayguade.
\newblock {A Systematic Methodology to Generate Decomposable and Responsive
  Power Models for {CMPs}}.
\newblock \emph{IEEE Trans. Comput.}, 62\penalty0 (7):\penalty0 1289--1302,
  July 2013.
\newblock ISSN 0018-9340.
\newblock \doi{10.1109/TC.2012.97}.
\newblock URL \url{http://dx.doi.org/10.1109/TC.2012.97}.

\bibitem[Brouwers et~al.(2014)Brouwers, Zuniga, and
  Langendoen]{Brouwers:2014:NNE:2668332.2668337}
N.~Brouwers, M.~Zuniga, and K.~Langendoen.
\newblock {NEAT: A Novel Energy Analysis Toolkit for Free-Roaming Smartphones}.
\newblock In \emph{Proceedings of the 12th ACM Conference on Embedded Network
  Sensor Systems}, SenSys '14, pages 16--30, New York, NY, USA, 2014. ACM.
\newblock ISBN 978-1-4503-3143-2.
\newblock \doi{10.1145/2668332.2668337}.
\newblock URL \url{http://doi.acm.org/10.1145/2668332.2668337}.

\bibitem[Cao et~al.(2012)Cao, Blackburn, Gao, and
  McKinley]{Cao:2012:YYP:2337159.2337185}
T.~Cao, S.~M. Blackburn, T.~Gao, and K.~S. McKinley.
\newblock {The Yin and Yang of Power and Performance for Asymmetric Hardware
  and Managed Software}.
\newblock In \emph{Proceedings of the 39th Annual International Symposium on
  Computer Architecture}, ISCA '12, pages 225--236, Washington, DC, USA, 2012.
  IEEE Computer Society.
\newblock ISBN 978-1-4503-1642-2.
\newblock URL \url{http://dl.acm.org/citation.cfm?id=2337159.2337185}.

\bibitem[Casella and Berger(1990)]{CasBer90}
G.~Casella and R.~Berger.
\newblock \emph{{Statistical Inference}}.
\newblock Duxbury Press Belmont, Calif, 1990.

\bibitem[Chang et~al.(2003)Chang, Farkas, and Ranganathan]{compaq_energy}
F.~Chang, K.~I. Farkas, and P.~Ranganathan.
\newblock {Energy-Driven Statistical Sampling: {Detecting} Software Hotspots}.
\newblock In \emph{Proceedings of the 2nd International Conference on
  Power-Aware Computer Systems}, PACS'02, pages 110--129, Berlin, Heidelberg,
  2003. Springer-Verlag.
\newblock ISBN 3-540-01028-9.
\newblock URL \url{http://dl.acm.org/citation.cfm?id=1766991.1767002}.

\bibitem[Contreras and Martonosi(2005)]{prediction_profiling1}
G.~Contreras and M.~Martonosi.
\newblock {Power Prediction for {Intel} {XScale} Processors Using Performance
  Monitoring Unit Events}.
\newblock In \emph{Proceedings of the 2005 International Symposium on Low Power
  Electronics and Design}, ISLPED '05, pages 221--226, New York, NY, USA, 2005.
  ACM.
\newblock ISBN 1-59593-137-6.
\newblock \doi{10.1145/1077603.1077657}.
\newblock URL \url{http://doi.acm.org/10.1145/1077603.1077657}.

\bibitem[Curtis-Maury et~al.(2008)Curtis-Maury, Shah, Blagojevic, Nikolopoulos,
  de~Supinski, and Schulz]{dimitrios_bronis}
M.~Curtis-Maury, A.~Shah, F.~Blagojevic, D.~S. Nikolopoulos, B.~R. de~Supinski,
  and M.~Schulz.
\newblock {Prediction Models for Multi-dimensional Power-Performance
  Optimization on Many Cores}.
\newblock In \emph{Proceedings of the 17th International Conference on Parallel
  Architectures and Compilation Techniques}, PACT '08, pages 250--259, New
  York, NY, USA, 2008. ACM.
\newblock ISBN 978-1-60558-282-5.
\newblock \doi{10.1145/1454115.1454151}.
\newblock URL \url{http://doi.acm.org/10.1145/1454115.1454151}.

\bibitem[Flinn and Satyanarayanan(1999{\natexlab{a}})]{powerscope99}
J.~Flinn and M.~Satyanarayanan.
\newblock {{PowerScope}: {A} Tool for Profiling the Energy Usage of Mobile
  Applications}.
\newblock In \emph{2nd IEEE Workshop on Mobile Computing Systems and
  Applications}, February 1999{\natexlab{a}}.

\bibitem[Flinn and Satyanarayanan(1999{\natexlab{b}})]{powerscope_flinn}
J.~Flinn and M.~Satyanarayanan.
\newblock {Energy-Aware Adaptation for Mobile Applications}.
\newblock In \emph{Proceedings of the Seventeenth ACM Symposium on Operating
  Systems Principles}, SOSP '99, pages 48--63, New York, NY, USA,
  1999{\natexlab{b}}. ACM.
\newblock ISBN 1-58113-140-2.
\newblock \doi{10.1145/319151.319155}.
\newblock URL \url{http://doi.acm.org/10.1145/319151.319155}.

\bibitem[Ge et~al.(2010)Ge, Feng, Song, Chang, Li, and Cameron]{PowerPack}
R.~Ge, X.~Feng, S.~Song, H.-C. Chang, D.~Li, and K.~W. Cameron.
\newblock {{PowerPack}: {Energy} Profiling and Analysis of High-Performance
  Systems and Applications}.
\newblock \emph{IEEE Transactions on Parallel and Distributed Systems},
  21\penalty0 (5):\penalty0 658--671, 2010.
\newblock ISSN 1045-9219.
\newblock URL \url{http://doi.ieeecomputersociety.org/10.1109/TPDS.2009.76}.

\bibitem[Hsu et~al.(2012)Hsu, Kuehn, and Poole]{Hsu:2012:TES:2188286.2188309}
C.-H. Hsu, J.~A. Kuehn, and S.~W. Poole.
\newblock {Towards Efficient Supercomputing: {Searching} for the Right
  Efficiency Metric}.
\newblock In \emph{Proceedings of the 3rd ACM/SPEC International Conference on
  Performance Engineering}, ICPE '12, pages 157--162, New York, NY, USA, 2012.
  ACM.
\newblock ISBN 978-1-4503-1202-8.
\newblock \doi{10.1145/2188286.2188309}.
\newblock URL \url{http://doi.acm.org/10.1145/2188286.2188309}.

\bibitem[{Intel Corporation}()]{vtune}
{Intel Corporation}.
\newblock {{Intel}\textsuperscript{\textregistered} Performance Tuning
  Utility}.
\newblock URL \url{https://software.intel.com/en-us/ intel-vtune-amplifier-xe}.

\bibitem[{Intel Corporation}(2009)]{intel_manual_rapl}
{Intel Corporation}.
\newblock \emph{{{Intel}\textsuperscript{\textregistered} 64 and IA-32
  Architectures Software Developer's Manual}}.
\newblock Number 253669-033US. December 2009.

\bibitem[Isci and Martonosi(2003)]{prediction_profiling2}
C.~Isci and M.~Martonosi.
\newblock {Runtime Power Monitoring in High-End Processors: {Methodology} and
  Empirical Data}.
\newblock In \emph{Proceedings of the 36th Annual IEEE/ACM International
  Symposium on Microarchitecture}, MICRO 36, pages 93--, Washington, DC, USA,
  2003. IEEE Computer Society.
\newblock ISBN 0-7695-2043-X.
\newblock URL \url{http://dl.acm.org/citation.cfm?id=956417.956567}.

\bibitem[Kansal and Zhao(2008)]{energy_profiling_microsoft}
A.~Kansal and F.~Zhao.
\newblock {Fine-Grained Energy Profiling for Power-Aware Application Design}.
\newblock \emph{SIGMETRICS Perform. Eval. Rev.}, 36\penalty0 (2):\penalty0
  26--31, Aug. 2008.
\newblock ISSN 0163-5999.
\newblock \doi{10.1145/1453175.1453180}.
\newblock URL \url{http://doi.acm.org/10.1145/1453175.1453180}.

\bibitem[Keranidis et~al.(2014)Keranidis, Kazdaridis, Passas, Igoumenos,
  Korakis, Koutsopoulos, and Tassiulas]{Keranidis:2014:NMM:2602044.2602047}
S.~Keranidis, G.~Kazdaridis, V.~Passas, G.~Igoumenos, T.~Korakis,
  I.~Koutsopoulos, and L.~Tassiulas.
\newblock {{NITOS} Mobile Monitoring Solution: {Realistic} Energy Consumption
  Profiling of Mobile Devices}.
\newblock In \emph{Proceedings of the 5th International Conference on Future
  Energy Systems}, e-Energy '14, pages 219--220, New York, NY, USA, 2014. ACM.
\newblock ISBN 978-1-4503-2819-7.
\newblock \doi{10.1145/2602044.2602047}.
\newblock URL \url{http://doi.acm.org/10.1145/2602044.2602047}.

\bibitem[Li et~al.(2013)Li, de~Supinski, Schulz, Nikolopoulos, and
  Cameron]{Li:2013:SER:2420628.2420808}
D.~Li, B.~R. de~Supinski, M.~Schulz, D.~S. Nikolopoulos, and K.~W. Cameron.
\newblock {Strategies for Energy-Efficient Resource Management of Hybrid
  Programming Models}.
\newblock \emph{IEEE Trans. Parallel Distrib. Syst.}, 24\penalty0 (1):\penalty0
  144--157, Jan. 2013.
\newblock ISSN 1045-9219.
\newblock \doi{10.1109/TPDS.2012.95}.
\newblock URL \url{http://dx.doi.org/10.1109/TPDS.2012.95}.

\bibitem[Lohr(1999)]{lohr:1999}
S.~Lohr.
\newblock \emph{{Sampling: Design and Analysis}}.
\newblock Brooks/Cole, 1999.

\bibitem[Manousakis et~al.(2014)Manousakis, Zakkak, Pratikakis, and
  Nikolopoulos]{Manousakis2014}
I.~Manousakis, F.~S. Zakkak, P.~Pratikakis, and D.~S. Nikolopoulos.
\newblock {{TProf}: {An} Energy Profiler for Task-Parallel Programs}.
\newblock \emph{Sustainable Computing: Informatics and Systems}, 2014.
\newblock ISSN 2210-5379.
\newblock \doi{http://dx.doi.org/10.1016/j.suscom.2014.07.004}.
\newblock URL \url{http://www.sciencedirect.com/ science/ article/ pii/
  S2210537914000390}.

\bibitem[McIntire et~al.(2007)McIntire, Stathopoulos, and
  Kaiser]{McIntire:2007:ESN:1236360.1236448}
D.~McIntire, T.~Stathopoulos, and W.~Kaiser.
\newblock {Etop: {Sensor} Network Application Energy Profiling on the {LEAP2}
  Platform}.
\newblock In \emph{Proceedings of the 6th International Conference on
  Information Processing in Sensor Networks}, IPSN '07, pages 576--577, New
  York, NY, USA, 2007. ACM.
\newblock ISBN 978-1-59593-638-7.
\newblock \doi{10.1145/1236360.1236448}.
\newblock URL \url{http://doi.acm.org/10.1145/1236360.1236448}.

\bibitem[Montgomery and Runger(2002)]{montgomery2002applied}
D.~Montgomery and G.~Runger.
\newblock \emph{Applied Statistics and Probability for Engineers}.
\newblock Wiley, 2002.
\newblock ISBN 9780471204541.
\newblock URL \url{http://books.google.co.uk/books?id=w8WYQgAACAAJ}.

\bibitem[{ODROID}()]{odroid}
{ODROID}.
\newblock {ODROID-XU+E}.
\newblock URL \url{http://hardkernel.com/main/products/}.

\bibitem[Pathak et~al.(2012)Pathak, Hu, and Zhang]{eprof_mobile}
A.~Pathak, Y.~C. Hu, and M.~Zhang.
\newblock {Where is the Energy Spent Inside My App?: {Fine} Grained Energy
  Accounting on Smartphones with Eprof}.
\newblock In \emph{Proceedings of the 7th ACM European Conference on Computer
  Systems}, EuroSys '12, pages 29--42, New York, NY, USA, 2012. ACM.
\newblock ISBN 978-1-4503-1223-3.
\newblock \doi{10.1145/2168836.2168841}.
\newblock URL \url{http://doi.acm.org/10.1145/2168836.2168841}.

\bibitem[Ren et~al.(2010)Ren, Tune, Moseley, Shi, Rus, and Hundt]{gwp}
G.~Ren, E.~Tune, T.~Moseley, Y.~Shi, S.~Rus, and R.~Hundt.
\newblock {Google-Wide Profiling: {A} Continuous Profiling Infrastructure for
  Data Centers}.
\newblock \emph{IEEE Micro}, 30\penalty0 (4):\penalty0 65--79, July 2010.
\newblock ISSN 0272-1732.
\newblock \doi{10.1109/MM.2010.68}.
\newblock URL \url{http://dx.doi.org/10.1109/MM.2010.68}.

\bibitem[Schubert et~al.(2012)Schubert, Kostic, Zwaenepoel, and Shin]{eprof}
S.~Schubert, D.~Kostic, W.~Zwaenepoel, and K.~G. Shin.
\newblock {Profiling Software for Energy Consumption}.
\newblock \emph{2012 IEEE International Conference on Green Computing and
  Communications}, 0:\penalty0 515--522, 2012.
\newblock \doi{http://doi.ieeecomputersociety.org/10.1109/GreenCom.2012.86}.

\bibitem[Shao and Brooks(2013)]{Shao:2013:ECI:2648668.2648758}
Y.~S. Shao and D.~Brooks.
\newblock {Energy Characterization and Instruction-Level Energy Model of
  {Intel's} {Xeon} {Phi} Processor}.
\newblock In \emph{Proceedings of the 2013 International Symposium on Low Power
  Electronics and Design}, ISLPED '13, pages 389--394, Piscataway, NJ, USA,
  2013. IEEE Press.
\newblock ISBN 978-1-4799-1235-3.
\newblock URL \url{http://dl.acm.org/citation.cfm?id=2648668.2648758}.

\bibitem[Tallent et~al.(2009{\natexlab{a}})Tallent, Mellor-Crummey, Adhianto,
  Fagan, and Krentel]{hpctool_analysis}
N.~R. Tallent, J.~M. Mellor-Crummey, L.~Adhianto, M.~W. Fagan, and M.~Krentel.
\newblock {Diagnosing Performance Bottlenecks in Emerging Petascale
  Applications}.
\newblock In \emph{Proceedings of the Conference on High Performance Computing
  Networking, Storage and Analysis}, SC '09, pages 51:1--51:11, New York, NY,
  USA, 2009{\natexlab{a}}. ACM.
\newblock ISBN 978-1-60558-744-8.
\newblock \doi{10.1145/1654059.1654111}.
\newblock URL \url{http://doi.acm.org/10.1145/1654059.1654111}.

\bibitem[Tallent et~al.(2009{\natexlab{b}})Tallent, Mellor-Crummey, and
  Fagan]{hpdctool_unwind}
N.~R. Tallent, J.~M. Mellor-Crummey, and M.~W. Fagan.
\newblock {Binary Analysis for Measurement and Attribution of Program
  Performance}.
\newblock In \emph{Proceedings of the 2009 ACM SIGPLAN Conference on
  Programming Language Design and Implementation}, PLDI '09, pages 441--452,
  New York, NY, USA, 2009{\natexlab{b}}. ACM.
\newblock ISBN 978-1-60558-392-1.
\newblock \doi{10.1145/1542476.1542526}.
\newblock URL \url{http://doi.acm.org/10.1145/1542476.1542526}.

\bibitem[Tsoi and Luk(2011)]{Tsoi:2011:PPO:2082156.2082159}
K.~H. Tsoi and W.~Luk.
\newblock {Power Profiling and Optimization for Heterogeneous Multi-core
  Systems}.
\newblock \emph{SIGARCH Comput. Archit. News}, 39\penalty0 (4):\penalty0 8--13,
  Dec. 2011.
\newblock ISSN 0163-5964.
\newblock \doi{10.1145/2082156.2082159}.
\newblock URL \url{http://doi.acm.org/10.1145/2082156.2082159}.

\bibitem[Tu et~al.(2014)Tu, Hsu, Chen, Chen, and
  Hung]{Tu:2014:PPP:2597648.2566660}
C.-H. Tu, H.-H. Hsu, J.-H. Chen, C.-H. Chen, and S.-H. Hung.
\newblock {Performance and Power Profiling for Emulated Android Systems}.
\newblock \emph{ACM Trans. Des. Autom. Electron. Syst.}, 19\penalty0
  (2):\penalty0 10:1--10:25, Mar. 2014.
\newblock ISSN 1084-4309.
\newblock \doi{10.1145/2566660}.
\newblock URL \url{http://doi.acm.org/10.1145/2566660}.

\bibitem[Wilke et~al.(2013)Wilke, G\"{o}tz, and
  Richly]{Wilke:2013:JGF:2451605.2451610}
C.~Wilke, S.~G\"{o}tz, and S.~Richly.
\newblock {{JouleUnit}: {A} Generic Framework for Software Energy Profiling and
  Testing}.
\newblock In \emph{Proceedings of the 2013 Workshop on Green in/by Software
  Engineering}, GIBSE '13, pages 9--14, New York, NY, USA, 2013. ACM.
\newblock ISBN 978-1-4503-1866-2.
\newblock \doi{10.1145/2451605.2451610}.
\newblock URL \url{http://doi.acm.org/10.1145/2451605.2451610}.

\end{thebibliography}



\end{document}